\documentclass[namedreferences,hyperref,optionalrh,solaromanenum]{spr-sola}
\usepackage{microtype}
\usepackage{amsmath}
\usepackage{graphicx}                    
\usepackage{color}                       
\usepackage{lineno}                      

\chardef\us=`\_

\begin{document}
\nolinenumbers

\begin{frontmatter}

\title{Data-driven segmentation of solar filaments based on ground-based instrument data}

%
\author[addressref=aff1,corref,email={stepanyuk.oleg@aol.com}]{\inits{O.}\fnm{Oleg}~\snm{Stepanyuk}\orcid{}}
\author[addressref=aff2,email={werner.poetzi@uni-graz.at}]{\inits{W.}\fnm{Werner}~\snm{P\"otzi}\orcid{0000-0003-4811-0543}}
\author[addressref=aff1,email={mdechev@nao-rozhen.org}]{\inits{M.}\fnm{Momchil}~\snm{Dechev}\orcid{0000-0002-2049-4998}}
\author[addressref=aff1,email={rmiteva@nao-rozhen.org}]{\inits{R.}\fnm{Rositsa}~\snm{Miteva}\orcid{0000-0002-0938-5678}}
\author[addressref=aff1,email={kkozarev@nao-rozhen.org}]{\inits{K.}\fnm{Kamen}~\snm{Kozarev}\orcid{0000-0002-6591-4482}}

\address[id=aff1]{Institute of Astronomy and National Astronomical Observatory (IANAO), Bulgarian Academy of Sciences, 1784 Sofia, Bulgaria}
\address[id=aff2]{Institute of Physics, Kanzelh\"ohe Observatory for Solar and Environmental Research, University of Graz, 8010 Graz, Austria}

%
\runningauthor{Stepanyuk et al.}
\runningtitle{Solar filament segmentation and tracking}

\begin{abstract}

The dynamic behavior of solar prominences and filaments is a precursor to coronal mass ejections (CMEs), which can disrupt Earth's magnetosphere and affect satellite communications. Systematic ground-based solar observations, conducted with high temporal resolution, are instrumental in monitoring these structures. Analysis of the morphological changes and destabilization processes of filaments and prominences captured in datasets can help identify early warning signs of potential eruptions. This capability is vital for developing reliable space weather forecasting systems, thereby mitigating the adverse effects of solar disturbances on Earth's technological infrastructure. Previously, we introduced hybrid algorithmic-data driven method which performance was shown on a set of Solar Dynamic Observatory (SDO) Atmospheric Imaging Assembly (AIA) instrument data performing segmentation of extreme ultraviolet (EUV) and shock waves. In this work, we extend certain aspects of this approach to image segmentation of solar features -- filaments and prominences using H\ensuremath{\alpha} Kanzelh\"ohe Observatory data.  We describe a leakage-aware, sparse multi-year sampling strategy. We present image segmentation models pre-trained on synoptic and normalized image data, followed by a discussion of how model performance depends on training-set engineering, data preprocessing techniques, and the choice of loss metrics. We perform a limited study, both analytically and numerically, of the effect of external, fixed normalization versus native input for compact U-Net-based segmentation of solar H\ensuremath{\alpha} filaments, also contributing to ongoing research on normalization in deep networks in general and compact models in particular.

\end{abstract}

%

\end{frontmatter}

%
\section{Introduction}\label{s:Intro} 

Solar prominences/filaments are isolated structures which constitute of dense, cool plasma suspended in the hot corona by magnetic fields \citep{2014LRSP...11....1P}. They are observed at multiple wavelengths, including H\ensuremath{\alpha}, extreme ultraviolet (EUV), and radio emissions \citep{Labrosse:2010, Vourlidas:2013}. In H\ensuremath{\alpha}, they appear as bright, elongated structures when seen above the limb and are called "prominences", while on the disk, they manifest as dark, filamentary features due to absorption \citep{Engvold:2015, TandbergHanssen:1995} and the term "filaments" is used.

Visually, eruptive filaments (EFs) are significantly smaller, typically around 20 Mm (megameters) in length, while non-eruptive filaments can extend up to 200 Mm. Another key visual distinction is the presence of a pre-eruptive EUV sigmoid -- a bright, S-shaped structure visible in EUV observations -- that often appears in EFs but is rarely seen in their non-eruptive counterparts. These visual cues provide initial hints about a filament’s eruptive potential.

Among the two types, eruptive prominences (EPs) are particularly significant due to their dynamic behavior and association with solar eruptive phenomena, such as coronal mass ejections (CMEs) and solar flares \citep{2003ApJ...586..562G,2013AdSpR..51.1967S}. These structures can remain stable for days or weeks before suddenly erupting, releasing vast amounts of plasma and magnetic energy into the heliosphere \citep{2007SoPh..245..287G}.

The underlying difference between the two types stems from their magnetic structures and the energy they store \citep{2010SSRv..151..333M}. EFs are supported by magnetic flux ropes (MFRs) with high magnetic free energy, which is the excess energy in the magnetic field beyond its lowest stable state. This elevated energy makes EFs prone to instability, often leading to eruptions. In contrast, non-eruptive filaments have MFRs with lower magnetic free energy, enhancing their stability. Additionally, the height of the MFR plays a critical role: EFs reach heights where the overlying magnetic field weakens, triggering an eruption via the torus instability. Non-eruptive filaments, however, remain at lower heights, where the stronger overlying field keeps them confined \citep{2015PASJ...67...78J}.

The height-time and velocity-time profiles of EFs suggest a strong correlation with underlying magnetic field gradients, supporting torus instability models \citep{2020ApJ...894...85C}. Moreover, the prominence mass appears to influence the terminal speed of the eruption, with higher-mass EFs achieving lower final velocities \citep{2017ApJ...846...86B}.

Eruptions often begin with a slow rise phase, followed by a rapid acceleration that can exceed the local Alfvén speed \citep{2022ApJ...926...18W}. This acceleration is typically associated with the onset of a CME, where the prominence material is carried outward by the expanding magnetic structure \citep{2012LRSP....9....3W}. Spectroscopic observations reveal plasma motions, including rotations, shearing flows, and helical structures, suggesting the presence of twisted magnetic flux ropes \citep{2014ApJ...797...52Y}. Initiation of the eruption may originate from mechanisms such as tether-cutting reconnection \citep{1989ApJ...343..971V}, flux cancellation \citep{2003ApJ...585.1073A}, breakout reconnection \citep{1999ApJ...510..485A}, flux emergence \citep{2000ApJ...545..524C}, or mass-loading \citep{2011ApJ...727L..10S}, as well as ideal MHD instabilities such as kink \citep{2005ApJ...630L..97T} and torus instability \citep{2006PhRvL..96y5002K}.

Magnetically, EPs are frequently associated with coronal cavities, i.e., low-density, elliptical regions observed above the solar limb in EUV and white-light images \citep{2006ApJ...641..590G,2013SoPh..288..603F}. These cavities are often interpreted as the cross-sections of coronal flux ropes and serve as precursors to eruption. Observations indicate that cavities undergo slow inflation prior to eruption, possibly due to helicity accumulation or flux emergence at the photosphere \citep{2011A&A...533L...1R}. In recent years, statistical and kinematic analyses have provided new constraints on the eruption onset conditions. Studies have identified a critical height range ($\sim$0.13--0.19 $R_{\odot}$) beyond which prominences become gravitationally and magnetically unstable \citep{2001JGR...10625177F,2012ApJ...744..168L}. High-resolution observations from space-based missions such as the Solar and Heliospheric Observatory (SOHO), Hinode and the Solar Dynamics Observatory (SDO) have provided unprecedented details on their thermodynamic and kinematic properties \citep{2012ApJ...758L..37B,2014ApJ...794..149C}.

Significant part of CMEs are associated with EPs, which are often observable in coronagraphs as dense cores within three-part CME structures. These associations emphasize the critical role of EPs in the initiation and morphology of CMEs \citep{1979phsp.coll..252R,2010SSRv..151..333M}, which are major drivers of geomagnetic storms \citep{1999SSRv...88..529G}. The entrained cold plasma can be detected in situ as "flux rope" signatures in solar wind data \citep{2010ApJ...723L..22L}. Thus, understanding their eruption mechanisms is also critical for space weather forecasting \citep{2021LRSP...18....4T}.

Ground-based and L1 observations complement magnetogram data and other solar imaging techniques, feeding into predictive models that help estimate the likelihood, velocity, and geoeffectiveness of CMEs. When such eruptions head toward Earth, the resulting interplanetary CME (ICME) can cause geomagnetic storms that in turn can interfere with satellite communications and navigation (e.g., the global positioning system, GPS), and ground-based infrastructure (electrical grids, pipelines). As a result, tracking filament (prominence) evolution is a vital component of space weather forecasting, offering a practical means of mitigating risks to both technological infrastructure and human activities reliant on stable space-based services.

The study of EPs is crucial for understanding solar activity, space weather, and the underlying magnetohydrodynamic (MHD) processes governing their evolution \citep{2000JGR...10523153F,2011LRSP....8....6S}. Routine observation of these structures can be valuable for CME forecasting. In particular, filaments (prominences) can signal the buildup of magnetic stress, highlight regions of potential instability, and provide critical insights into the direction and strength of any ensuing eruptions. By examining their shape, size, and overall behavior, we gain the ability to identify potential solar disturbances well in advance, often hours or even days before they occur. Future missions like Solar Orbiter \citep{2020A&A...642A...1M} and Parker Solar Probe will provide multi-point observations, improving our understanding of eruptive prominences in the heliosphere \citep{2015ASSL..415.....V}, thus putting even more demands for development of better processing and tracking techniques.  

In recent years machine learning (ML) and deep learning (DL) methods have become more frequently applied to specific solar physics problems, as they are well suited to extracting useful information from time series and large multi-wavelength imaging data. For instance, \citet{Szenicer:2019} used a convolutional neural network to produce EUV irradiance maps from Atmospheric Imaging Assembly (AIA) images. \citet{Li:2013} proposed a multi-layer model to predict solar flares based on sequential sunspot data; and \citet{Kim:2019} applied generative adversarial networks (GAN) to generate the magnetic flux distribution of the Sun from SDO/AIA images. \citet{Nishizuka:2018} tested three different ML algorithms -- support vector machine (SVM), k-nearest neighbors (k-NN), and multi-layer perceptron (MLP) -- for predicting solar flare occurrences within the next 24 hours. 

Encoder--decoder architectures for semantic segmentation, such as U-Net, offer structural advantages over generic object detectors when the target task requires pixel-level delineation rather than bounding boxes. U-Net’s symmetric encoder–decoder with skip connections directly produces dense masks and was designed to train from scratch on limited annotated data using strong augmentation, rather than relying on large-scale pretraining on datasets like ImageNet or COCO. A typical U-Net has on the order of 31M parameters, but recent work shows that channels and depth can be aggressively reduced to ~2M parameters and ~3.4B Multiply-Accumulate Operations (MACs)  (16× fewer MACs than the original) with only modest degradation in segmentation accuracy \citep[][and PyTorch implementations]{posso_embeddedunet:2025}. 
In many biomedical and remote-sensing benchmarks, U-Net and its variants routinely obtain Dice scores around 0.9 while remaining relatively compact (e.g., DCSSGA-UNet: \citealp{hussain:2025}).
In contrast, modern one-stage detectors such as YOLO11, EfficientDet, SSD and RTMDet are optimized for bounding-box detection mAP on datasets like COCO. YOLO11 small, for example, achieves about 47\% mAP@50-95 on COCO with a compact architecture and runs at tens of milliseconds per frame on cloud-class GPUs. EfficientDet-D0 attains 34.6\% mAP with only 3.9M parameters but relies on a BiFPN and multi-scale feature fusion tailored to detection. \citep{tan_efficientdet:2020} RTMDet-tiny pushes the speed-accuracy frontier further, reaching ~41\% Average Precision(AP) at over 1000 FPS with ~4.8M parameters on an RTX 3090 \citep{lyu_rtmdet:2022}, While some of these detectors have segmentation extensions, their primary design target remains box-level detection, and they typically require non-maximum suppression and more complex post-processing than a pure segmentation network.


Automated H-alpha filament detection was first formalized as a computational problem in the early 2000s. Shih and Kowalski (2003) introduced an automatic extraction and tracking method
for H-alpha filaments based on two alternative preprocessing techniques, followed by intensity thresholding and morphological filtering to identify candidate filament pixels and trace their temporal evolution across serial disk images. This work established the foundational pipeline-disk finding, background correction, thresholding, morphological cleaning, and spine extraction that most subsequent classical methods would follow \citep{Shih:2003}. Qu et al. (2005) presented a five-stage algorithm applied to BBSO full-disk H-alpha images that remains one of the most cited early works in the field. Their pipeline combined the Stabilized Inverse Diffusion Equation (SIDE) for contrast enhancement and contour sharpening, adaptive local thresholding for filament extraction, a Support Vector Machine (SVM) classifier to discriminate filaments from sunspots, and morphological thinning operations to extract filament spines, footpoints, and detect filament disappearances \citep{Qu:2005}. \citet{Yuan:2011} developed a generic segmentation and characterization method applicable to full-disk images from multiple observatories — BBSO, Mauna Loa, and others — without requiring observatory-specific recalibration. Their approach introduced a cascading Hough circle detector for solar disk localization, polynomial surface fitting for luminance correction, adaptive thresholding, and graph-theory-based spine identification.


\citet{Ahmadzadeh:2019} applied Mask R-CNN — which in a way could be thought as the heavyweight predecessor of YOLO-style detectors — to BBSO/HEK filament data, establishing that deep detection networks can scale with additional annotations and handle multiple solar feature types in a unified model, though pixel-mask precision remained limited. Further advance came with the work by \citet{Diercke:2024}, who trained YOLOv5 (large configuration, no pretrained weights) on manually labelled ChroTel images, then used its bounding-box predictions on GONG images as pseudo-labels to train a U-Net segmentation model at a scale infeasible with manual annotation alone. YOLOv5 achieved mean accuracy 84.8\%, precision 74.5\%, recall 84.9\% on ChroTel; the U-Net trained on pseudo-labels reached mean accuracy 92.4\% and recall 93.2\% on the ChroTel test set archive, enabling pixel-level physical parameter extraction (area, tilt angle, latitudinal position) sufficient to reconstruct the Solar Cycle 24 rush-to-the-pole. The two-step practice proposed by Diercke et al. (2024) --- detection first for cheap scalable labelling, segmentation second for scientific measurement --- conceptually has similarities with the approach earlier proposed by \citet{Stepanyuk:2022,Stepanyuk:2024,Stepanyuk:2025} for detection of Coronal Bright Fronts (CBF) from SDO/AIA and coronagraph data. 
The practical lesson from this body of work is architectural complementarity: Earlier YOLO models attempted for solar imagery segmentation and detection-first models remain unmatched for throughput and label-efficient detection of large, high-contrast filaments, but they are structurally unable to produce the pixel-wise masks required for morphological science. The fundamental limitation of earlier YOLO-class and Mask R-CNN frameworks in the filament segmentation context stems from the mismatch between their primary task — bounding-box prediction — and the elongated, irregular morphology of filaments. Filaments routinely span tens of degrees of heliographic longitude as thin, curved threads, violating the roughly isotropic, spatially compact object assumption implicit in anchor-based detection grids
\citep{Hussain:2024}.
More fundamentally, a bounding box cannot deliver filament area, tilt angle, or length; extracting those requires a secondary thresholding step that reintroduces sunspot misclassifications. The architectural ceiling is not a tuning problem—it is structural.

Generally, Large pre-trained backbones (Mask R-CNN's ResNet, YOLO's feature-pyramid neck) were designed for natural-image datasets such as MS-COCO, where dozens of object categories coexist against cluttered backgrounds. In H-alpha imagery the background is photometrically stable and the target has a single, well-defined absorption character. \citet{Zhu:2025} state explicitly, ``the relatively uniform and stable background of solar images actually reduces the overall complexity of the task''. Carrying tens of millions of parameters tuned for background diversity is therefore not an asset but a liability — manifesting as overfitting, representational overhead, and memory budgets incompatible with potential running on on-board instrument hardware, while raising solar physics community interest in compact models. U-Net's encoder-decoder topology with symmetric skip connections was designed for precisely the regime that H-alpha filament imagery represents: dense pixel prediction over images where the
target occupies a minority of pixels, context from multiple spatial scales is needed to delineate boundaries, and annotated data are scarce \citep{ronneberger_unet:2015}. \citet{Zhu:2019} provided the first end-to-end application of U-NET to BBSO full-disk H-alpha images. They added dropout layers at 50\% neuron-drop probability to counter overfitting on the small training set and replaced transposed convolution with nearest-neighbour 
interpolation in the expansive path, converging in less than an hour on a single GPU. The model attained DSC 0.8944 and TPR 0.9145 on ten held-out test images, with segmentation accuracy scaling positively with filament size.  \citet{Zhu:2025} introduced Flat U-Net, an ultralightweight model keeping channel depth constant across all encoder and decoder layers — eliminating the quadratic parameter growth of the standard doubling-channel design. Channel interactions are modelled through Simplified Channel Attention (SCA) and Channel Self-Attention (CSA) blocks. Applied to raw BBSO FITS images without limb-darkening correction, Flat-fielding, or cloud-cover masking, Flat U-Net achieved DSC 0.82 and recall 0.74 with only 0.26 million parameters.
Large-scale detection architectures such as YOLOv5 and Mask R-CNN were optimised for natural image datasets containing dozens of object categories with complex, cluttered backgrounds; their large parameter counts and feature pyramid necks are engineered for that diversity, but the same capacity becomes a liability — expressed as overfitting, representational overhead, and inflated  memory requirements — when applied to a domain where the background is photometrically predictable and the target class has a single, well-defined absorption character.

Distinct dimension of the argument for U-Net-class over YOLO-class models becomes apparent when deployment constraints are considered. The YOLOv5 large configuration used by Diercke et al. (2024) carries approximately 46 million parameters \citep{Jocher:2020}; the YOLOv5x variant reaches approximately 87 million \citep{Jocher:2020}. These parameter counts are manageable on well-equipped research workstations but incompatible with the embedded compute environments of next-generation space-borne solar instruments, which require models to run onboard with strict memory and power budgets. The standard U-Net ([64, …, 1024] channel configuration) itself is not lightweight — but the U-Net family is structurally better positioned for radical parameter
reduction than detection architectures, because the encoder-decoder topology does not require a multi-scale region-proposal backbone.

Recent theoretical and empirical studies have examined the effects of input normalization on deep networks. \citet{Ioffe:2015} formalized batch normalization as a reversible, learnable affine transform; \citet{Salimans:2016} showed that the principal optimization benefits of normalization arise from weight-side reparameterization rather than from the inputs themselves; \citet{Yu:2022} demonstrated that outer-layer scaling is especially sensitive to fixed preprocessing; \citet{Shao:2020} proved that deep networks can match or exceed batch-normalized baselines even without any normalization layers; and \citet{Kociolek:2020} provided medical-imaging evidence that aggressive per-image normalization can degrade feature stability. However, this body of work is developed for generic deep classifiers and for natural- or medical-image benchmarks; it does not address compact convolutional encoder–decoder architectures such as U-Net, it does not treat pixel-level semantic segmentation of sparse, photometrically defined targets, and — critically — it has not been examined in the solar-physics domain, where the input photometry itself (limb darkening, instrument response, solar-cycle phase) carries physical signal.

\citet{Stepanyuk:2024} presented algorithmic multi-instrument solar eruptive feature recognition and tracking methods and applied them to tracking of coronal bright fronts and CMEs from the low corona out to 30 solar radii, using ground- (COSMO K-Coronagraph) and space-based (SDO/AIA, SOHO/LASCO C2 and C3) telescopic observations.Natural evolution of this approach was application of Machine Learning (ML) techniques towards segmentation of high-dynamic range heliophysics observations. \citet{Stepanyuk:2025} trained Convolutional Neural Network (CNN) image segmentation models for SDO/AIA instrument data using feature masks obtained from the Wavetrack code followed by discussion about the optimal network architecture, tracking and velocity estimation techniques.

The purpose of this work is to extend the segmentation, training-set engineering, and parameter-tuning aspects of the approach presented in \citep{Stepanyuk:2025} to data-driven segmentation of on-disk solar features—specifically prominences and filaments—using Kanzelhöhe Observatory data \citep{Potzi:2014}, and to further study the role of input data preprocessing in deep segmentation networks

The present work is, to our knowledge, if the first limited study of the effect of external, fixed normalization versus native synoptic input specifically for compact U-Net segmentation in general and of solar H\ensuremath{\alpha} filaments in particular, and to do so both analytically and numerically (training-dynamics metrics and held-out performance). This is the gap that separates our contribution from prior studies and constitutes its novelty.


The paper is organized as follows. Section~\ref{method_sec} describes the methodological framework, including the U-Net-based model architecture together with the metrics and loss functions chosen for class-imbalanced solar filament segmentation. Section~\ref{input_sec} presents the Kanzelh\"ohe Solar Observatory H\ensuremath{\alpha} input data and details our approach to training-set engineering using real synoptic/normalized observations and rotation-based data augmentation, followed by initial propositions about methodological choices and the role of input data normalization in the current research and in deep segmentation networks in general. Section~\ref{results_sec} reports the model training routine and quantitative segmentation performance analysis. Within subsections ~\ref{discussion_norm_subsec} and~\ref{unet_norm_subsec} we give the analytical explanation of these results, discussing internal, learned normalization inside the network. The same section discusses prospects for compact models in embedded applications such as CCOR and SDO/AIA. In section ~\ref{limitations_sec} we describe limitations of the current research and Section  ~\ref{conclusions_sec} summarizes the main findings and outlines directions for future work. 

\section{The Method}
\label{method_sec}

Image segmentation is a fundamental technique in computer vision that divides an image into multiple segments or regions based on visual characteristics such as texture, intensity, or other attributes. The goal is to produce a representation of the image that highlights semantically meaningful components, making further analysis more effective. The output typically consists of labeled regions that correspond to distinct parts of the image. This process involves assigning a class label to each pixel; consequently, if multiple objects of the same class are present, they are treated as a single entity in the segmentation map. In contrast, models that identify and differentiate between individual instances of the same class are referred to as instance segmentation models.

\subsection{Model Architecture}

\begin{figure}[htp]
\centering
\includegraphics[width=12cm]{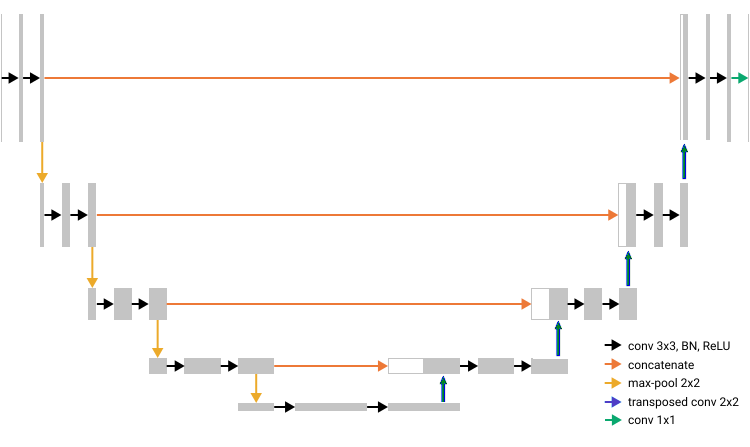}
\caption{U-NET architecture in general. Each gray box corresponds to a multi-channel feature map. (The number of channels is not specified, as it would depend on the network setup in each specific case) }
\label{fig:fig_unet_fig_1}
\end{figure}

In \citep{Stepanyuk:2025} authors have searched for a DL architecture, which would perform well for low contrast, high noise data such as SDO AIA instrument data, and at the same time require a relatively small size of the training set. 
Our methodology is likewise grounded in the U-Net architecture \citep{ronneberger_unet:2015}, which features a contracting path succeeded by an expansive path, linked through a central bottleneck (Fig. \ref{fig:fig_unet_fig_1}). This efficient and effective framework has facilitated its widespread application across diverse fields, ranging from satellite imagery analysis to industrial defect detection. In the contracting or encoder path, the network commences with an input image and implements a sequence of convolutional layers, each succeeded by a rectified linear unit (ReLU) activation. These convolutional operations are followed by max-pooling layers that down-sample the image. This procedure effectively encapsulates the contextual information of the image while concurrently diminishing its spatial dimensions, though augmenting its depth, defined as the number of feature maps. The central component of the U-Net, termed the bottleneck, comprises two $3\times3$ convolutions, each followed by a ReLU layer, without any ensuing max-pooling. This layer resides between the contracting and expansive paths and encodes the most compressed representation of the input image. The expansive or decoder path of the network up-samples the feature maps.

At each stage of this path, the network up-samples the feature map, executes a $2\times2$ up-convolution to enlarge its spatial dimensions, and subsequently concatenates it with a correspondingly cropped feature map from the contracting path. Following this concatenation, two $3\times3$ convolutions followed by ReLU activations are applied. These linkages from the contracting path, commonly designated as ``skip connections,'' enable the network to preserve spatial information that could otherwise be forfeited during the down-sampling process in the encoder. The architecture concludes with a final $1\times1$ convolution layer that projects each feature vector to the requisite number of classes, yielding the segmented output. A prominent attribute of U-Net is its skip connections that interconnect the encoder and decoder paths. These connections are essential for retaining finer details, which are critical for accurate segmentation tasks. An additional benefit of the U-Net architecture lies in its comparatively fewer parameters relative to other analogous architectures. This characteristic not only accelerates training but also renders it especially apt for scenarios involving limited training data availability. Our overarching approach to CNN design is not rigidly constrained to a particular CNN architecture such as U-Net, which was selected for the present study, and permits the investigation of alternative architectures in future work (see hyper-parameter search).
For the segmentation of larger images, U-Net utilizes an overlap-tile strategy, thereby ensuring that even the border regions of images, where data may occasionally be absent, are predicted with accuracy.

Optical H\ensuremath{\alpha} instrument data, due to its lower dynamic range, may be less challenging and may not impose such strict requirements on the network architecture. Additionally, it is generally easier to engineer large and diverse training sets for this type of data. Nevertheless, the shape of the segmented features—i.e., solar filaments, despite being relatively more distinct than CBF's—places specific demands on the thoughtful selection of appropriate loss metrics.

\subsection{Metrics. Loss function}

Classical performance measures such as pixel-wise accuracy or standard loss functions like mean squared error and cross-entropy are often inadequate for image segmentation tasks characterized by severe class imbalance and fine-scale structures. In solar image segmentation, foreground features such as filaments typically occupy only a small fraction of the total image area, causing accuracy to be dominated by the correctly classified background and rendering it insensitive to meaningful segmentation errors. Similarly, conventional losses tend to favor majority classes, allowing trivial solutions (e.g., predicting only background) to achieve deceptively high scores while failing to capture thin or elongated structures. As a result, overlap-based metrics that directly quantify spatial agreement between predicted and ground-truth regions—such as the Dice coefficient and Intersection over Union (IoU)—are better suited for both evaluation and optimization in such contexts.
The Intersection over Union (IoU), also known as the Jaccard index, and the Dice coefficient (also known as the Sørensen–Dice coefficient) are the most used metrics to quantify the similarity between two images. 

IoU is defined as the size of the intersection divided by the size of the union of two sets. In the context of comparing two images or segmentation masks, it is calculated as:

\[
\mathrm{IoU}(A, B) = \frac{|A \cap B|}{|A \cup B|}
\]

The Dice coefficient is similar to IoU but it gives more weight to the intersection part. It is defined as twice the size of the intersection divided by the sum of the sizes of the two sets:

\[
\mathrm{Dice}(A, B) = \frac{2 \times |A \cap B|}{|A| + |B|}
\]

Dice coefficient is better suited for thin image features like solar filaments because it places greater emphasis on the intersection (overlap) relative to the individual set sizes, making it more sensitive to small structures and overlaps. This sensitivity helps in accurately capturing fine details without harshly penalizing minor discrepancies in thin, low-area regions, where even small errors could drastically reduce scores in other metrics. In contrast, IoU tends to yield lower values for the same overlap due to its inclusion of the full union in the denominator, which can undervalue precise matches in sparse or thin features.

For thicker features like CBFs (Coronal Bright Fronts or similar dense structures in solar imaging), IoU is more appropriate as it provides a balanced assessment by equally considering both intersection and the non-overlapping parts (union), penalizing under- or over-segmentation more stringently. This makes IoU better for larger, bulkier areas where overall spatial agreement is crucial, and errors in broader regions need stronger correction without overemphasizing just the overlap.

Dice coefficient is particularly effective for calculating accuracy and loss during U-Net training for filament segmentation because it inherently handles class imbalance well—filaments occupy minimal pixel area compared to background, so Dice focuses on optimizing overlap rather than being skewed by dominant classes. As a differentiable metric (often used as Dice loss = 1 - Dice), it enables smooth gradient-based optimization, leading to better convergence on thin, intricate structures. Additionally, its sensitivity to small overlaps encourages the model to refine boundaries and details in sparse features like filaments, improving generalization in solar image segmentation tasks.

Specifically, to further illustrate the advantages of the Dice coefficient for thin image features like solar filaments, one can consider the mathematical relationship between Dice and IoU: 
\[
\mathrm{Dice} = \frac{2 \times \mathrm{IoU}}{1 + \mathrm{IoU}}
\]
and 
\[
\mathrm{IoU} = \frac{\mathrm{Dice}}{2 - \mathrm{Dice}}.
\]

This transformation shows that Dice consistently yields higher values than IoU for the same overlap, with the amplification most pronounced when $\mathrm{IoU}$ is small (i.e., sparse overlaps typical in thin structures). For instance, in the limit of small intersection where $|A \cap B| \ll |A|, |B|$, $\mathrm{IoU} \approx \frac{|A \cap B|}{|A| + |B|}$, while $\mathrm{Dice} \approx \frac{2 |A \cap B|}{|A| + |B|}$, making Dice roughly twice IoU in this sparse-overlap limit. The relative
sensitivity of the two metrics to a fixed reduction in intersection can
then be compared explicitly.
If a small perturbation $\delta$ reduces the intersection while the individual set sizes remain approximately constant, the relative drop in the Dice coefficient is simply 
\[
\frac{\Delta D}{D} = \frac{\delta}{|A \cap B|}.
\]

To see why, let $I = |A \cap B|$. Since $D = \frac{2I}{|A|+|B|}$, a reduction $I \to I - \delta$ gives $\Delta D = \frac{2\delta}{|A|+|B|}$, and hence $\frac{\Delta D}{D} = \frac{\delta}{I}$.

For the IoU, $J = \frac{I}{|A|+|B|-I}$. Let $U = |A|+|B|-I$. Under the same perturbation, $J \to \frac{I-\delta}{U+\delta}$. The relative drop is
\[
\frac{\Delta J}{J} = \frac{\delta (|A|+|B|)}{I (|A|+|B|-I+\delta)}.
\]
Because $|A|+|B| > |A|+|B|-I+\delta$ (for $I > \delta$), we have
\[
\frac{\Delta J}{J} > \frac{\delta}{I} = \frac{\Delta D}{D}.
\]

Consequently, under this perturbation model, the relative change in IoU is strictly larger than that of Dice. Thus, for a fixed reduction in overlap while the individual set sizes remain approximately constant, Dice exhibits a smaller relative decrease than IoU. This provides a simple mathematical illustration of why Dice can be advantageous for evaluating sparse, thin structures such as solar filaments.



For thicker features like CBFs, IoU's formulation provides a stricter balance by incorporating the full union, which penalizes discrepancies in larger areas more effectively. In regimes where $|A \cap B|$ is comparable to $|A|$ and $|B|$ (dense overlaps), the relation $\mathrm{IoU} = \frac{\mathrm{Dice}}{2 - \mathrm{Dice}}$ compresses high Dice values downward, making IoU more responsive to over- or under-segmentation. For example, near perfect overlap ($\mathrm{Dice} \to 1$), a small decrease $\epsilon$ in Dice causes IoU to drop by approximately 2$\frac{\epsilon}{ (2 - \mathrm{Dice})^2 }$ (via derivative), which is larger than Dice's linear drop, ensuring stronger emphasis on overall spatial agreement in bulky structures where union discrepancies dominate.

The Dice coefficient's effectiveness as a loss function in U-Net training for filament segmentation stems from its differentiable form, typically 
\[
\mathrm{Dice\ loss} = 1 - \frac{2 \sum_i p_i g_i + \epsilon}{\sum_i p_i + \sum_i g_i + \epsilon},
\] 
where $p_i \in [0,1]$ is the predicted probability, $g_i \in \{0,1\}$ is the ground truth, and $\epsilon > 0$ prevents instability. 
The gradient of the Dice loss is global, since each pixel prediction contributes to the overlap and normalization terms. Defining
\begin{equation}
N = 2\sum_j p_j g_j + \epsilon,
\qquad
D = \sum_j p_j+\sum_j g_j+\epsilon,
\end{equation}
the Dice loss is
\begin{equation}
L_{\mathrm{Dice}}=1-\frac{N}{D}.
\end{equation}
Applying the quotient rule gives
\begin{equation}
\frac{\partial L_{\mathrm{Dice}}}{\partial p_i}
=
\frac{N-2g_iD}{D^2},
\end{equation}
or explicitly,
\begin{equation}
\boxed{
\frac{\partial L_{\mathrm{Dice}}}{\partial p_i}
=
\frac{
2\sum_j p_jg_j+\epsilon
-
2g_i
\left(
\sum_jp_j+\sum_jg_j+\epsilon
\right)
}{
\left(
\sum_jp_j+\sum_jg_j+\epsilon
\right)^2
}.
}
\end{equation}


In contrast, IoU loss approximations (e.g., via relaxations like $\sum_i \max(p_i, g_i) \approx \sum_i p_i + g_i - p_i g_i$) yield less smooth gradients, potentially hindering convergence on thin features due to the union's non-local nature.

\section{Input data. Training sets engineering}
\label{input_sec}

Kanzelhöhe Solar Observatory (KSO), located near Villach, Austria, conducts ground-based observations of the Sun with emphasis on H\ensuremath{\alpha} and white-light imaging. These observations offer near real-time data and an online archive that captures photospheric and chromospheric features, including sunspots, filaments, prominences, and flare activity. The observatory’s high-cadence data collection facilitates continuous tracking of these solar phenomena and supports both short-term monitoring and long-term research into solar variability also contributing data to international networks like the Global H\ensuremath{\alpha} Network and ESA's Space Situational Awareness program 
\citep{ESA_KSO:2025}

The observatory archives H-alpha full-disk images in two main branches with distinct processing levels (Pötzi et al. and official KSO data products)\cite{Potzi:2014}.
Synoptic images are the daily "normal-contrast" overview products (one best-quality image per observing day). The solar disk is centred, a heliographic grid is overlaid, Solar North is rotated upward, and mild unsharp masking is applied for slight contrast enhancement. Natural center-to-limb variation(limb darkening / CLV) is retained, so intensity falls off toward the limb exactly as observed. These are intended for quick daily visual inspection and long-term synoptic monitoring.
Normalized (high-contrast) images undergo additional post-processing on top of the synoptic level. Global anisotropies and center-to-limb variation are removed by flat-fielding and division by a quiet-Sun intensity profile derived from the CLV correction table. The result is a photometrically uniform disk with homogeneous contrast across the entire field, making faint filaments, prominences, and fine structures far more visible.

The processing levels and sampling choices described in this section define the input regimes and training-set construction that are subsequently examined, both numerically (Sect.~\ref{results_sec}) and analytically (Sects.~\ref{discussion_norm_subsec} and~\ref{unet_norm_subsec}). We present them as the part of the methodological setup on which those later analyses depend, including one of the central questions of this study—the role of input data normalization in deep segmentation networks.


We randomly selected two datasets—each containing roughly 800 image–pair samples from 2021, 2022, and 2023—for both normalized and synoptic H\ensuremath{\alpha} imagery. 
When running training routine to segment solar filaments, one of the most insidious — and most common — methodological errors is the per-image random train/validation split.
Every standard machine-learning splitting recipe --- rests on a single statistical assumption: that samples are independent and identically distributed. The textbook justification for cross-validation, generalization-error bounds, and even the meaning of "validation Dice" all collapse without it. 
Solar filament data can violate this assumption decisively. A quiescent filament observed by the observatory typically persists on the disk for one to two solar rotations -- days to weeks. During that lifetime, the system records dozens to hundreds of frames of the same filament, separated by minutes. Between consecutive frames, the filament barely moves. Solar rotation carries it across the disk at roughly 13° per day, which is about half a degree per hour. The shape evolves on timescales of hours to days. The surrounding quiet-Sun pattern is essentially frozen on a per-day basis. The instrumental flat-field and seeing class are constant within an observing run. The consequence: pixel-IoU between consecutive masks of the same filament can be very high. The frames are not independent samples — they are near-duplicates of the same physical structure.
In such case the model encodes similar pixel pattern of recurring scenes. Validation samples sit at near-zero distance from their nearest training neighbors in input space.

In domains where this leakage mode has been quantified — medical imaging is the most-studied — the gap between per-image-random splitting and properly disjoint splitting typically falls in the range: Dice 0.10–0.25 for large recurrent objects (radiology benchmarking work, \citep{Bussola:2020}), Dice 0.05–0.15 for small or transient objects.

We follow sparse multi-year sampling targeting the worst kind of leakage — adjacent-minute frames of the same filament. Even two frames a day apart (let alone four days apart) are no longer pixel-near-duplicates of each other. The background pattern has shifted, the filament has moved several degrees in longitude due to solar rotation, the seeing conditions are different, the flat-field calibration may differ. That removes the dense-temporal-correlation problem. Solar Cycle 25 timeline includes 2021 -- rising phase (~30 spots/month), 2022 -- strong ascent, (80 spots/month), 2023 -- near maximum, (120 spots/month), 2024 -- actual maximum, (150+ spots/month). Thus 2021/2022/2023 sample is not a sample from a stationary distribution — it samples three quite different physical regimes: Filament population is different (more active-region filaments in 2023, more quiescent in 2021), Class prevalence (foreground pixel fraction) is different, background solar atmosphere/chromospheric context is different.

On this temporally de-correlated sample, each base dataset of H\ensuremath{\alpha} image–mask pairs (2021–2023) was partitioned, before augmentation, into train/validation/test subsets in a standard 70/15/15 ratio, stratified by observing year so that 2021, 2022, and 2023 are represented in in each subset. After applying data-augmentation techniques, the training sets expanded to around 4,000 training pairs. The purpose of data augmentation with rotation is to help the model learn to recognize objects from different viewpoints and orientations, which can improve its ability to generalize and make accurate predictions on new, unseen data. It helps to reduce over-fitting, as the model is exposed to more diverse examples of the same object. For that purpose, we used 45 degree rotation step of every image-mask pair from the initial dataset for training of all of our models presented in this publication.

The choices set out in this section define the first component of an argument about input normalization in compact segmentation networks. The synoptic-versus-normalized distinction mentioned here is approached numerically in Sect.~\ref{results_sec} and explained analytically in Sects.~\ref{discussion_norm_subsec}--\ref{unet_norm_subsec}, where we discuss how synoptic processing already performs a physics-aware normalization, that internal batch/group normalization performs a learned and reversible one, and that an additional external, fixed per-image normalization could be redundant and, for sparse and photometrically defined targets such as H\ensuremath{\alpha} filaments, potentially detrimental. The same definitions later underpin the generalizability argument of Sect.~\ref{compact_models_potential_subsec}, where possibility of extending of the approach to instruments such as CCOR and SDO/AIA is discussed. 
The models are updated periodically, please consult the relevant model description/README on GitLab for the most up-to-date details on architecture, training-set parameters, and evaluation metrics, as certain parameters may change over time.

\section{The Results}

\label{results_sec}

\subsection{Model training routine and performance on image segmentation tasks}
\label{training_subsec}

The baseline architecture (64 filters in the first layer, 3$\times$3 filter size, 4 layers depth, 0.1 standard deviation pixel intensity augmentation) after adjustment of depth, width and filters size/count based on models performance (see hyper-parameter search)  \citep{Stepanyuk:2025}. To optimize the training duration without compromising the model's ability to generalize, we implement an early stopping technique. Early stopping involves monitoring a chosen performance metric on a validation set at each epoch and stopping the training when this metric stops improving or begins to degrade.

In H\ensuremath{\alpha} imagery from observatories like KSO, solar flares appear as abrupt intensity enhancements, often manifesting as bright ribbons or patches within active regions associated with sunspot groups. These brightenings result from the rapid heating of chromospheric plasma during magnetic reconnection events. Filaments, dense plasma structures suspended along magnetic neutral lines, are observed as dark, elongated absorptions against the brighter chromospheric disk due to their cooler temperatures (around 5,000--8,000 K) causing H\ensuremath{\alpha} line absorption; when erupting, they may destabilize, leading to partial or full disappearance and associated brightenings. Prominences, the limb-view counterparts of filaments, emit brightly against the dark coronal backdrop, appearing as towering, loop-like or hedgerow-shaped structures that can exhibit dynamic surging, twisting, or ejection during eruptive phases. 

Normalization is typically applied to standardize input data, reduce effects like limb darkening or instrumental variations, and often intended to make training more consistent for machine learning. Our results show that models trained on synoptic data outperform the ones trained on normalized H\ensuremath{\alpha} data . 

\begin{figure}[htp]
\centering
\includegraphics[width=12cm]{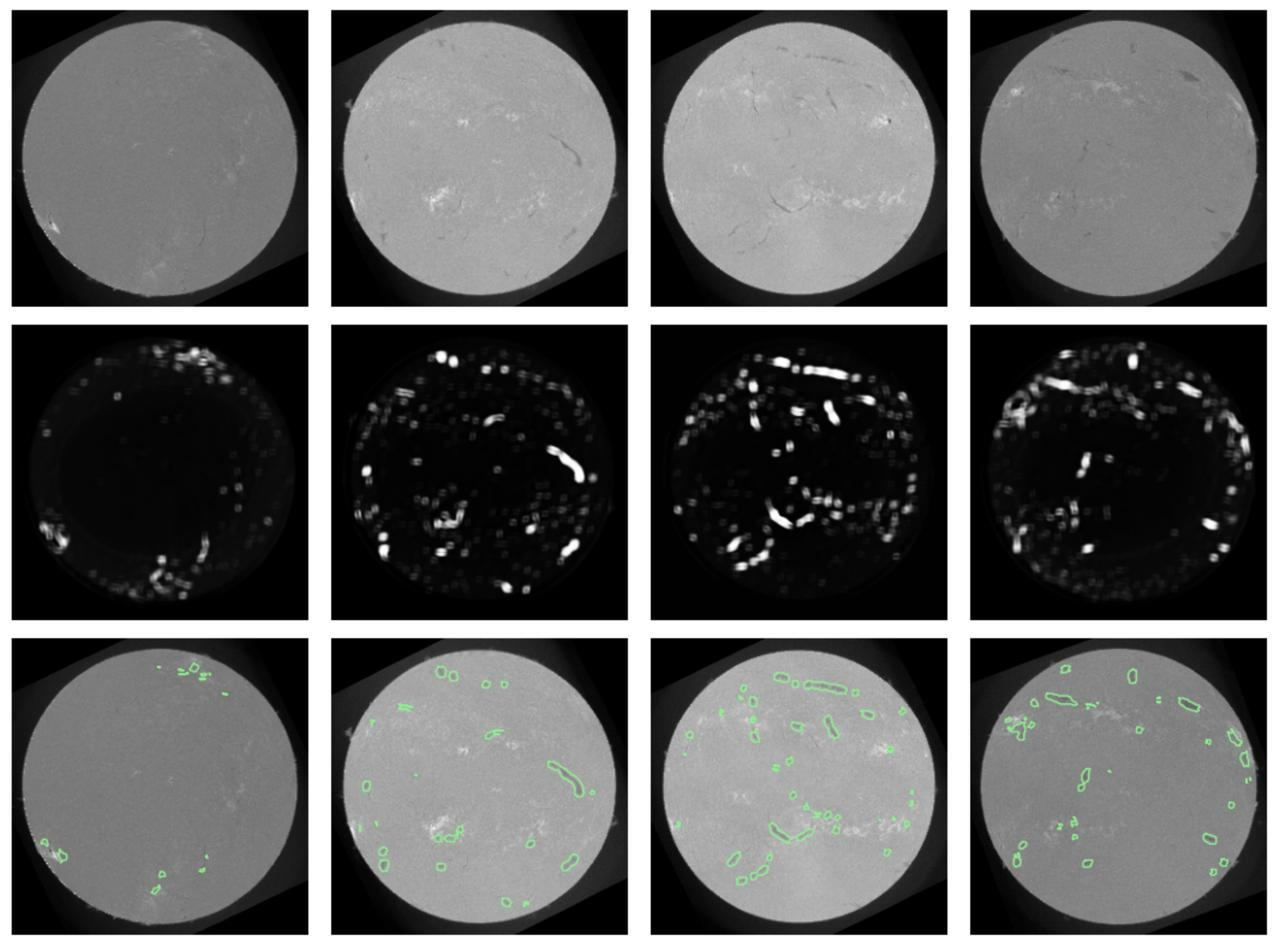}
\caption{U-Net image segmentation outputs for a model trained and evaluated on normalized H\ensuremath{\alpha} solar data. Top row: Input normalized H\ensuremath{\alpha} full-disk images of the Sun, showing faint dark linear features against a uniform gray background. Bottom row: Corresponding binary segmentation masks, where white pixels indicate detected solar features (filaments on a black background)}
\label{fig:halpha_normalized}
\end{figure}

\begin{figure}[htp]
\centering
\includegraphics[width=12cm]{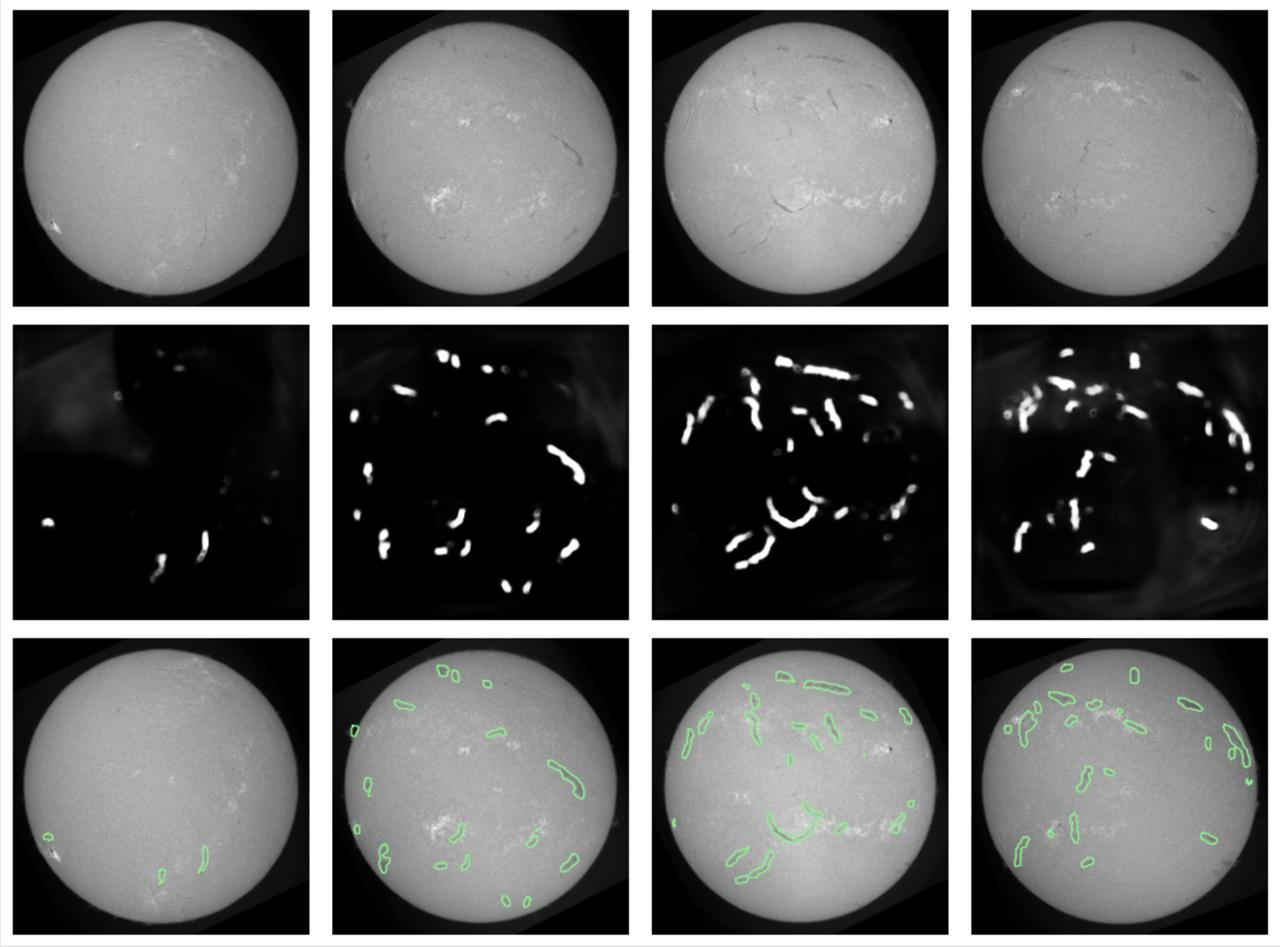}
\caption{U-Net image segmentation outputs for a model trained and evaluated on synoptic solar data. Top row: Input synoptic full-disk images of the Sun in grayscale, displaying slightly more pronounced linear features than normalized counterparts. Bottom row: Corresponding binary segmentation masks, highlighting detected features (distinctively filaments) in white on black, with improved continuity and detail}
\label{fig:halpha_synoptic}
\end{figure}

A simple intuitive explanation, which naturally suggests itself in this case it that normalization reduces the dynamic range and edge sharpness of features, blending them into the background and making it harder for the CNN (which relies on convolutional layers to detect local patterns and textures) to learn discriminative representations during training. For instance, filaments appear as narrow absorption regions with gradual intensity transitions; normalization might flatten these gradients, leading to under-segmentation or fragmentation, as seen in the masks of the figure \ref{fig:halpha_normalized}. Synoptic data, by retaining unnormalized variations, provides richer textural cues and higher contrast, enabling the model to generalize better and produce more accurate, connected masks. 
Overall, while normalization aids consistency, it can sometimes strip away valuable signal for tasks like segmentation where raw physical contrasts are key. 
In the section \ref{discussion_sec} we provide a more rigorous explanation based on discussion which took place in ML community in the recent years. 

\subsection{Training Metrics} 
\label{training_metrics_sec}

The figure summarizes training and validation dynamics for two deep learning models segmenting solar filaments in H\ensuremath{\alpha} images: a synoptic model (top row) and a normalized model (bottom row). Each row displays the Dice coefficient over training epochs and the corresponding Dice loss.

\begin{figure}[htp]
\centering
\includegraphics[width=12cm]{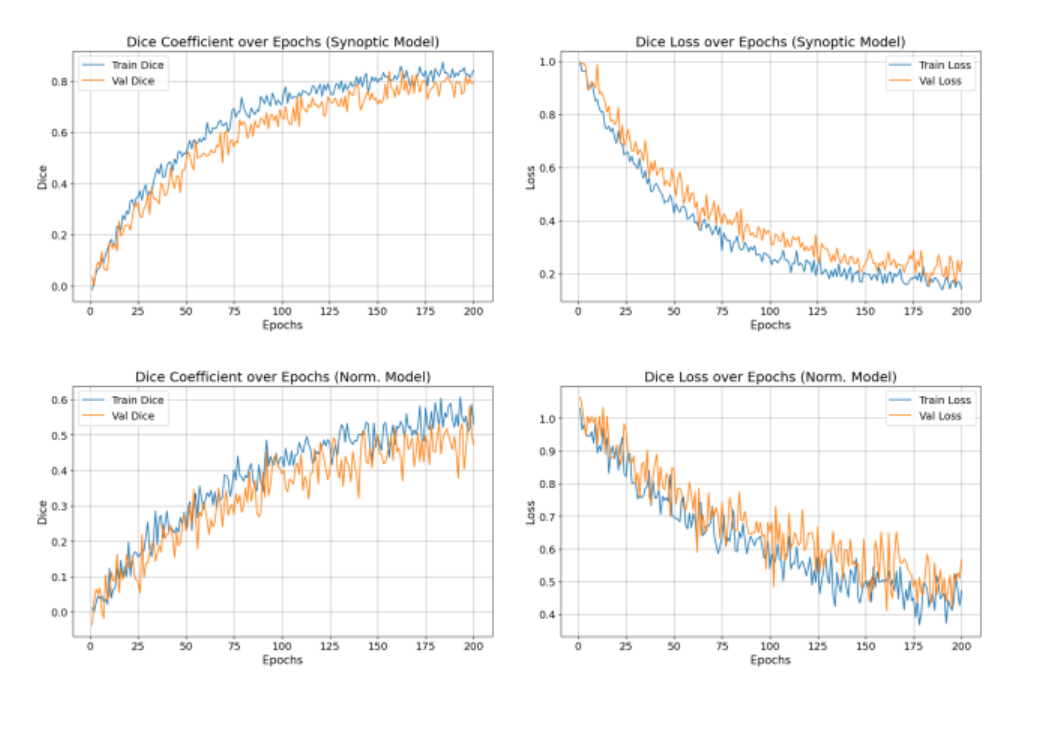}
\caption{Training/validation dynamics for two segmentation models. Top row: Synoptic model; bottom row: Normalized model. Columns (left→right): Dice coefficient over epochs (left) and Dice loss over epochs (right)}
\label{fig:metrics}
\end{figure}

In the synoptic model plots, the training (blue) and validation (orange) Dice coefficients rise smoothly and stabilize by 200 epochs at approximately 0.80--0.90 (train) and $\sim$0.75--0.80 (val), indicating strong learning with a small generalization gap. The Dice loss decreases monotonically to $\sim$0.20 on validation, matching the Dice trajectory and suggesting stable convergence without pronounced overfitting.

For the normalized model, the Dice coefficients increase more slowly and saturate at $\sim$0.50--0.60 on validation, with a noticeably larger train–validation gap. The corresponding Dice loss plateaus at a higher level ($\sim$0.45--0.55), consistent with weaker masks and less stable convergence.

Overall, the synoptic model’s higher Dice and lower loss across training and validation are compatible with the qualitative segmentations: predicted masks adhere more closely to filament structures with fewer spurious responses than those produced by the normalized model, which appear fragmented and prone to false positives in low-contrast regions.

Synoptic model metrics align with ranges reported in deep learning literature for similar tasks. For instance, an Attention U²-Net model reports Dice of 0.8323 and mIoU of 0.7139. An improved U-Net attains an average Dice of 0.8944. More sophisticated hybrid CNN-Transformer model achieves a Dice (F1) score of 0.9119 on H\ensuremath{\alpha} filament segmentation
The synoptic model's higher values roughly fall within these benchmarks for optimized models, while the normalized model's lower scores are consistent with baseline or suboptimal variants, highlighting the observed performance gap on our test dataset. 

\subsection{Discussion}
\label{discussion_sec}

\subsubsection{The conceptual Normalization inside the network, and \emph{external, fixed} normalization }
\label{discussion_norm_subsec}

Intuition that ``more normalization should help'' rests on a confusion between two operations that look alike 
but behave very differently: \emph{internal, learned} normalization inside the network, and \emph{external, fixed} normalization applied in the data loader. 
Within recent years, this subject provoked discussions in the Machine Learning community.

Ioffe and Szegedy~\citep{Ioffe:2015} define Batch Normalization (BN) as a sequence of three steps. Given mini-batch activations $\{x_1,\dots,x_m\}$ on one channel, BN computes the empirical mean and variance,
\begin{equation}
\mu_\mathcal{B} = \frac{1}{m}\sum_{i=1}^m x_i, \qquad
\sigma_\mathcal{B}^2 = \frac{1}{m}\sum_{i=1}^m (x_i - \mu_\mathcal{B})^2,
\label{eq:bn-stats}
\end{equation}
standardizes them,
\begin{equation}
\hat{x}_i = \frac{x_i - \mu_\mathcal{B}}{\sqrt{\sigma_\mathcal{B}^2 + \epsilon}},
\label{eq:bn-std}
\end{equation}
and applies a \emph{learnable} affine restoration
\begin{equation}
y_i = \gamma\,\hat{x}_i + \beta.
\label{eq:bn-affine}
\end{equation}

For a fixed mini-batch one can formally recover the original activations by setting
\[
\gamma = \sqrt{\sigma_B^2 + \epsilon}, \qquad \beta = \mu_B,
\]
which yields \(y_i = x_i\). This shows that the affine parameters that follow standardization have sufficient expressive power to compensate for the normalization. In practice, however, \(\gamma\) and \(\beta\) are learned by gradient descent and are not forced to invert the batch statistics; at inference time running averages are used instead of the current mini-batch statistics. The important distinction is therefore not strict reversibility or information preservation, but \emph{learnability and network context}: an external per-image normalization such as
\[
T(x) = \frac{x - \mu_x}{\sigma_x}
\]
is a fixed, non-learnable transformation applied before the data enter the network, so any information it removes is permanently lost from the network's point of view. In contrast, Batch Normalization is an internal operation whose subsequent affine transform is optimized end-to-end together with the rest of the model.


Salimans and Kingma \citep{Salimans:2016} sharpen this by showing that the optimization benefit normalization provides --- high learning rates, robust initialization, fast convergence --- comes from the \emph{weights}, not the inputs. Their reparameterization
\begin{equation}
\mathbf{w} = \frac{g}{\|\mathbf{v}\|}\,\mathbf{v}
\label{eq:wn}
\end{equation}
decouples weight magnitude from direction and recovers most of BN's training speed without touching the data at all. The familiar argument ``we must pre-normalize the inputs because otherwise the optimizer struggles'' is therefore weak: weight-side conditioning solves the same problem with no information loss on the input side.

Yu and Spiliopoulos \citep{Yu:2022} generalize the picture by parameterizing each layer's scaling as
\begin{equation}
h^{(i)}(x) = \frac{1}{N_i^{\gamma_i}} \sum_{j=1}^{N_i} \phi(W_j^{(i)\top} x + b_j), \qquad \gamma_i \in [1/2, 1],
\label{eq:ys}
\end{equation}
and proving that the network's variance and test accuracy depend on these exponents, with the outer (input- and output-adjacent) layers being far more sensitive than inner ones. Pre-normalization is functionally a fixed change to the first layer's effective scaling, and the first layer is precisely the place where a wrong scaling hurts most. A learned $\gamma, \beta$ reaches the optimum; a fixed min--max or z-score does not.

Shao et al. \citep{Shao:2020} close the loop by showing that deep networks can be trained \emph{without any normalization layers} and still match or beat BN baselines on ImageNet, MS-COCO segmentation, Kinetics, and WMT. If even internal normalization is dispensable, external pre-normalization is certainly not mandatory --- and the burden of proof is on showing that it adds something rather than removes something.

Kocio{\l}ek, Strzelecki and Obuchowicz \citep{Kociolek:2020} provide the empirical complement on real medical imagery: across Brodatz textures, kidney DCE-MRI, shoulder T2-MRI, and cardiac CT, the optimal normalization is dataset-dependent, no method dominates, and aggressive normalization combined with intensity-resolution reduction \emph{degrades} feature stability. For modalities whose physics lives in absolute photometry --- and solar synoptic H-$\alpha$ is such a modality --- pre-normalization is not neutral preprocessing; it is information removal.

The synthesis is straightforward. 
Synoptic processing already applies the standard KSO calibration and
presentation procedures, while retaining the natural center-to-limb
intensity variation. The normalized product applies additional processing
that removes the center-to-limb variation through flat-fielding and
division by a quiet-Sun intensity profile.
Internal BN/GN layers perform learned, channel- or group-dependent
normalization within the network. For Batch Normalization, mini-batch
statistics are used during training and running estimates are used during
inference, followed by a learnable affine transformation.

A second, hand-chosen, per-image normalization between these two is redundant where it agrees with them and destructive where it disagrees. Synoptic data feeds the network with photometric variation that encodes instrument response, class prevalence and solar-cycle phase (depending on the dataset) --- variation a high-capacity model uses as a free contextual feature. Pre-normalization erases this variation before the model can read it.

\subsubsection{U-Net: the gap between learned and fixed normalization}
\label{unet_norm_subsec}

U-Net amplifies the gap between learned and fixed normalization for three structural reasons. Encoder-decoder depth means many "internal normalizations". 
Each encoder block computes its own $\mu_\mathcal{B}, \sigma_\mathcal{B}^2$ on its own feature maps. Pre-normalization affects only the first block; every subsequent block recomputes statistics from scratch. The supposed benefit is therefore confined to a single layer, while the cost --- destroyed photometric content --- propagates through the whole network. Skip connections distribute the input-side bias to every decoder resolution.
A U-Net's skip connections feed shallow encoder features directly into the decoder. Whatever distortion a per-image pre-norm injects at the input rides down each skip and reappears at the matching decoder level, where it cannot be undone by the few BN layers downstream of the skip. The deepest decoder block sees both a heavily normalized bottleneck stream and a pre-norm-distorted skip. Internal BN at the concatenation tries to equalize them, but the per-image bias of the pre-norm is not absorbable by per-batch $\gamma, \beta$. This is likely a U-Net-specific compounding effect that does not exist in plain classifiers, at least to our knowledge. Per-channel $\gamma, \beta$ is richer than the segmentation literature usually credits.
In a filament U-Net, some channels need to be photometric (line-core depth distinguishing optically thick filaments from shallow plage), some morphological (thread-like darkening regardless of intensity), some contextual (suppressing umbra-as-filament). The learned $\gamma, \beta$ assign each channel its own scaling --- the photometric channels approach the identity, the morphological channels exploit the standardization, the suppressors learn negative gain. A fixed pre-norm forces every channel to behave as if it were morphological, removing the photometric basis the line-core channels depend on. 
The decoder must infer the corresponding information from the remaining
features, which may be more difficult if the removed intensity variation
contains discriminative information.

The Yu--Spiliopoulos sensitivity argument lands twice in U-Net rather than once: at the first encoder block (where pre-normalization sits) and at the final $1\times 1$ conv that produces the per-pixel logit map. Both are outer-layer sites, both are scaling-sensitive, and both are damaged when the input-side dynamic range has been collapsed.

The empirical convergence of the field reflects this. Diercke et al. \cite{Diercke:2024} train a U-Net on calibrated ChroTel and GONG H-$\alpha$ with only minimal global processing and obtain $\sim$94\% accuracy on filament segmentation, generalizing across instruments without retraining. SARD~\cite{Sard:2025} does the analogous job on HMI magnetograms with only a fixed $\pm 650$\,G clip and reaches AP $= 0.946$. Both leave the dynamic range intact and let internal Conv--BN--SiLU blocks handle the rest.

\subsection{Compact models: potential for embedded applications}
\label{compact_models_potential_subsec}

Although the present study focuses on image segmentation using ground-based instrument data, a broader perspective on practical applications highlights the further potential of compact encoder–decoder networks such as U-Net. A very similar approach can be directly extended to on-board spaceborne coronagraph data, including instruments such as CCOR (Compact Coronagraph), a next-generation visible-light coronagraph designed to continuously image the solar corona and detect Earth-directed coronal mass ejections in real time. This demonstrates strong prospects for autonomous, on-board processing under strict telemetry and power constraints. Moreover, U-Net-based approaches have already been shown to be highly effective on significantly more complex, multi-channel, and wide–dynamic-range solar imaging data, such as observations from the SDO/AIA (Solar Dynamics Observatory / Atmospheric Imaging Assembly), further supporting the generalizability and robustness of compact segmentation architectures in heliophysics applications \citep{Stepanyuk:2025}.
For embedded applications, several recent studies demonstrate that carefully slimmed and quantized U-Net-style models map well onto low-power hardware. An “embeddable” U-Net for aerial semantic segmentation reduces parameters from 31M to 1.9M and MACs from 55B to 3.4B and achieves real-time throughput (~60–80 FPS) on Jetson-class GPUs while maintaining competitive IoU \citep{posso_embeddedunet:2025}.
FPGA implementations of lightweight U-Net further show real-time semantic segmentation with tight power budgets \citep{modiboyina_fpgaunet:2024}.
In parallel, benchmarking work on edge devices such as Raspberry Pi and Jetson Orin reports that detector families like YOLOv8, EfficientDet-Lite and SSD face a trade-off between latency, energy consumption and detection accuracy across different boards and precision settings 
\citep{alqahtani_benchmarking:2024}.
Taken together, these results suggest that when the primary requirement is fine-grained, pixel-accurate outputs under tight compute and power constraints, compact encoder–decoder networks like U-Net (and their lightweight variants) can provide a more natural and efficient fit than repurposed box detectors, especially on resource-limited embedded platforms.

\section{Limitations}
\label{limitations_sec}

Possible limitations are coming from this particular experiment design, specifically training set engineering.  Splitting across the full 2021–2023 pool means that training and validation sets come from the same distribution, i.e., the validation does not measure performance on the radically new solar conditions. It measures performance on "interpolated" solar conditions. This isn't leakage in the strict sense, rather distribution shift not being tested, and generalization issue. Carrington-rotation recurrence is not very likely but potentially could cause soft leakage.

\section{Conclusions}
\label{conclusions_sec}

In this work, we developed and evaluated a data-driven pipeline for segmentation of on-disk solar filaments in H\ensuremath{\alpha} imagery from the Kanzelh\"ohe Solar Observatory (KSO). Building on our earlier hybrid algorithmic--CNN approach for SDO/AIA EUV features and shock waves \citep{Stepanyuk:2025}, we adapted a U-Net-based encoder--decoder architecture and trained it on carefully engineered datasets of paired KSO image--mask samples augmented by rotations. The resulting models segment filaments and prominences directly from full-disk H\ensuremath{\alpha} frames, without hand-tuned thresholds.

One of the findings is that models trained on the observatory’s standard synoptic images consistently outperform those trained on the photometrically normalized high-contrast products. Quantitative evaluation on held-out test sets shows validation Dice coefficients  markedly higher than the range achieved by their normalized counterparts. This performance gap is explained by the interplay between external, fixed preprocessing and the internal, learned normalization mechanisms already embedded within the CNN. Synoptic images preserve the full photometric context---including instrument response, and natural intensity gradients---that the network exploits as discriminative features. In contrast, aggressive per-image normalization flattens these gradients, removes physically meaningful signal, and forces the model to reconstruct lost information from morphology alone, resulting in fragmented masks and reduced generalization. These results align with, and in some cases exceed, recent benchmarks reported for H\ensuremath{\alpha} filament segmentation while using significantly fewer parameters on a relatively compact training set, showing potential for embedded on-board applications for models with similar architecture and setup.

This segmentation capability enhances monitoring of filament evolution, providing early indicators of eruptive potential critical for space weather forecasting. By enabling systematic tracking of morphological changes, our approach can support the identification of precursors to CMEs, contributing to improved predictions of geomagnetic disturbances. 
The novelty of the present contribution lies not only in the application of our optimized U-Net pipeline to KSO H\ensuremath{\alpha} data but, more importantly, in the systematic analysis of how data-preprocessing choices propagate through the entire deep-learning pipeline. By contrasting synoptic and normalized inputs side-by-side---both analytically and through training-dynamics metrics and model performance evaluation---we provide physics-informed insight for the future solar-image segmentation efforts which future models, including larger architectures and cross-instrument training campaigns, can adopt to avoid unnecessary information loss and achieve reliable performance across heterogeneous archives.
As we study both analytically and numerically the effect of external, fixed normalization versus native  input for compact U-Net-based networks, we contribute to ongoing research on normalization in deep networks in general and compact models in particular.
Future improvements may include incorporating temporal sequences for dynamic tracking via recurrent or 3D CNNs, and optimizing for real-time embedded applications on other ground-based instruments. Additionally, exploring advanced architectures like Transformers or self-supervised learning could further boost performance on limited datasets, paving the way for more robust, automated solar feature analysis in operational space weather systems.



\subsection{Appendix. Practical suggestion for solar H-$\alpha$ U-Net segmentation} 
\label{appendix_sec}
One of the general conclusions coming from our analysis is that synoptic preprocessing supplies the physics-aware normalization, internal BN/GN supplies the learned,  per-channel normalization, and any external per-image normalization sitting between the two is at best redundant and at worst -- for sparse, photometrically defined classes like H-$\alpha$ filaments -- actively destructive.

We suggest that engineering future models with larger amount of parameters and/or cross-instrument training could benefit from the following setup constraints: 

\begin{itemize}
\item Keep synoptic photometric calibration intact; do not strip it with a per-image z-score or min--max.
\item Apply at most a fixed, dataset-global linear rescale to bring inputs into roughly $[-1,1]$ for optimizer stability. 


\item Place a normalization block immediately after each skip-connection concatenation in the decoder, so the encoder and bottleneck streams are equalized before the next conv.

\item For cross-instrument training (ChroTel + GONG + KSO), use 

histogram matching to a canonical reference frame rather than per-image equalization. This preserves relative photometry while removing instrument drift.

\item Leave the final $1\times 1$ conv un-normalized and use Dice or Tversky loss for class-imbalanced segmentation; let the internal $\gamma, \beta$ learn whatever per-channel scaling each feature map needs.
\end{itemize}

\begin{authorcontribution}
{OS created the model and performed the images segmentation and tracking, described large parts of the text and performed the final editing; WP provided the data, assisted in text writing and editing; MD contributed to the writing and editing of the text; RM contributed to the writing and coordinated the editing of the text; KK provided help in the model development and data visualizations.}
\end{authorcontribution}

\begin{fundinginformation}
This research was funded by the Bulgarian National Science Foundation project No. KP-06-Austria/5 (14-08-2023) and Austria’s Agency for Education and Internationalisation (OeAD) project No. BG 04/2023.
\end{fundinginformation}

\begin{dataavailability}
KSO data is publicly available at \url{https://www.kso.ac.at/index_en.php}.
Wavetrack and CNN models are publically available at 
\url{https://gitlab.com/iahelio/helios_cnn} and/or
\url{https://github.com/oleg-stepanyuk}

\end{dataavailability}

\begin{ethics}
\begin{conflict}
The authors declare no conflict of interest. 
\end{conflict}
\end{ethics}

\bibliographystyle{spr-mp-sola}
\bibliography{OS_bib}  

\end{document}